\documentstyle[psfig,preprint,aps]{revtex}
\def\beq{\begin{equation}}
\def\eeq{\end{equation}}
\def\bea{\begin{eqnarray}}
\def\eea{\end{eqnarray}}
\def\xx{\nonumber\\}
\def\w1{\omega_{k_0}}
\def\wq2{\omega_{k_0+q_0}}
\def\tww{\tilde{\omega}_{k_0}}
\def\twwp{\tilde{\omega}'_{k_0+q_0}}

\begin{document}

\date{\today}
\tighten

\title{A New Formulation of a $1+1$ Dimensional Field Theory Constrained to a
Box}

\author{ M.E. Carrington${}^{a,b}$, R. Kobes${}^{b,c}$ and G.
Kunstatter${}^{b,c}$}

\address{ ${}^a$ Department of Physics, Brandon University, Brandon, Manitoba,
R7A 6A9 Canada\\
${}^b$  Winnipeg Institute for
Theoretical Physics, Winnipeg, Manitoba\\
 ${}^c$ Department of Physics,
 University of Winnipeg, Winnipeg, Manitoba, R3B 2E9 Canada }

\maketitle

\begin{abstract}
We consider a $1+1$ dimensional field theory constrained to a finite box of
length $L$.  Traditionally, calculations in a box are done by replacing the
integrals over the spatial momenta by discrete sums and then
evaluating sums and doing analytic continuations.
We show that it is also possible to do such calculations using an analogy to
finite temperature field theory.  We develop a formalism that is similar to the
closed time path formulation of finite temperature field theory.  Our technique
 can be used to calculate spatially retarded green functions, without
evaluating sums or doing analytic continuations.  We calculate the self energy
in a simple scalar theory as an example.

\end{abstract}

\vspace{3ex}

\section{Introduction}

 We consider a $1+1$ dimensional field theory constrained to a finite box
in the spatial dimension.  We show that this system can be studied in two
different ways.  We can impose the boundary condition in the usual way by
restricting the momentum variable to take discrete values $k=2\pi n/L$ and
replacing the momentum integral by a sum over the discrete index.  We will
refer to this type of calculation as a `box' calculation.  We will show that we
can formulate this theory in a simpler way, based on an analogy with finite
temperature field theory. This approach will be called the `CSP' 
[Closed Space Path] formalism (the
notation will be explained below).  We study the self energy function for a
simple scalar theory and compare the results of the box calculation and the CSP
calculation.

The motivation is as follows. There are physical systems in which the physics
is essentially $1+1$ dimensional. One example is a micromaser, 
where one is interested
in the situation where a beam of excited atoms enters a box, interacts with a
radiation field which is trapped inside the box, and emerges from the other
end. The standard non-relativistic description of this system is the
Jaynes-Cummings model which describes a reservoir of two level atoms in
interaction with a single mode oscillator representing a coherent
electromagnetic field \cite{JC}.   It has been shown that this simple non-relativistic
model can be obtained as the non-relativistic limit of a relativistic quantum
field theory \cite{gmm}. The micromaser is studied experimentally by measuring
the probability that an atom will emerge in the excited state, given that it
entered in the excited state. It is known that this probability exhibits
oscillations, which are known as Rabi oscillations.  
Non-relativistically, it is
straightforward to obtain these oscillations from a calculation of transition
amplitudes using the Jaynes-Cummings Hamiltonian.  To obtain the relativistic
generalization of this result, we need to calculate the imaginary part of the
atomic self energy.  Measurements of the final atomic state are performed at a
specific point in space, and thus it is the spatially retarded self energy that
is relevant, and not the time retarded one. One can obtain this spatially
retarded self energy in a straightforward way using the CSP formalism.  The CSP
formalism is constructed in analogy to the closed time path formalism (CTP) of
finite temperature field theory \cite{Keld,Schw,Lifs} and has the same advantages with respect to the
box calculation as the CTP formalism has over the imaginary time formalism of
finite temperature field theory \cite{Joe,LeBellac,Henn}.  The CSP formalism allows us to obtain
spatially retarded green functions without explicitly
worrying about the difficulties associated with analytic continuation.

It is straightforward to see the similarity between a $1+1$
dimensional field theory constrained
to a box and finite temperature field theory.  Finite temperature field theory
involves expectation values of operators weighted with Boltzmann factors,
\bea \langle {\cal O}\rangle = \sum_n \langle \phi_n | e^{-\beta H} {\cal O} |
\phi_n\rangle \nonumber\eea
The Boltzmann factor is equivalent to an imaginary time evolution operator,
\bea \langle {\cal O}\rangle = \sum_n\langle \phi_n(x);t-i\beta|{\cal
O}|\phi_n(x);t\rangle \nonumber\eea
If the system obeys periodic boundary conditions such that
\bea \phi_n(t-i\beta) = \phi_n(t)\nonumber
\eea
then we have
\bea
\langle {\cal O}\rangle = \Sigma_n \langle \phi_n(t)|{\cal O}|\phi_n(t)\rangle
\nonumber
\eea
Thus, imposing a periodic boundary condition (in imaginary time) is equivalent
to taking expectation values that are weighted with a distribution function
$e^{-\beta H}$, which means that finite temperature field theory can be
formulated as a field theory constrained to a box in imaginary time.  This
constraint is imposed at the level of the action by using momentum expansions
of the fields with the Fourier integral over the zeroth momentum component
replaced by a sum over the discrete values $k_0 = 2\pi i n T$.  The generating
functional leads to Feynman rules which are the same as the zero temperature
rules except that the zero momentum components take discrete values, and the
part of the loop integral that corresponds to integrals over the zeroth
momentum components are replaced by summations. In the evaluation of Feynman
integrals, these sums are replaced by contour integrals that contain
distribution functions which have poles at the same discrete values 
$k_0 = 2\pi i n T$.  These contour integrals can be rotated so 
that the integration contour
encloses the real axis, running from negative infinity to positive infinity
above the real axis, and backwards from positive infinity to negative infinity
below the real axis.  This contour is called the closed time path contour.

We want to consider a $1+1$ dimensional field theory constrained to a box in real
space, instead of a box in imaginary time.  To achieve this, we will
define a field theory on a contour that encloses the real $x$-coordinate axis.
This closed space path (CSP) formalism will be mathematically very similar to
the CTP formalism of finite temperature field theory.
Consider what is involved in reversing the roles of space and time.
Relativistically, time and coordinate variables play similar roles.  The usual
formulation of a quantum field theory uses the fact that the system contains
both positive and negative energy states which move forward and backward in
time respectively. Propagators are constructed as symmetric linear combinations
of these states.  The conventional choice of energy eigenstates as basis states
gives a physical non-relativistic limit.  In the non-relativistic limit, the
positive and negative eigenstates are separated, and the non-relativistic
theory contains energy states of one sign only.  This restriction corresponds
to the fact that in the non-relativistic world, time runs only in the forward
direction, and physical states are positive energy states.
Relativistically, the special nature of time emerges only in statistical
situations.  When a system consists of many degrees of freedom which form a
heat bath, then the system will move towards equilibrium, and the forward time
direction is singled out as the physical direction.

Now consider our situation.  The CSP formalism is mathematically similar to the
CTP formalism with the roles of space and time reversed.  The rigorous
procedure for constructing the CSP formalism is presented in detail in
Section 3.  Propagators are constructed as symmetric combinations of positive
and negative momentum eigenstates, moving forward and backward in coordinate
space.  The imaginary time box of finite temperature field theory is replaced
by a box in real space, and the parameter $\beta=1/T$ which characterizes the
length of the box in imaginary time is replaced with a parameter $iL$ where $L$
is the length of the box in coordinate space.  The Boltzmann factor 
$e^{-\beta p_0}$ which gives the exponential decay that 
leads to equilibration, is
replaced by a factor $e^{-iLp}$ which gives rise to oscillatory behaviour.
Thus, in contrast to the finite temperature case, the system maintains its
ability to evolve in both spatial directions, even after statistical effects
have been included.

This paper is organized as follows.  We work with a simple scalar theory with a
Lagrangian of the form,
\bea {\cal L} = \frac{1}{2}[(\partial_t \phi_1)^2 - (\partial_x \phi_1)^2 -
m_1^2 ] + \frac{1}{2}[(\partial_t \phi_2)^2 - (\partial_x \phi_2)^2 - m_2^2
]-\frac{\lambda}{3!} (\phi_1+ \phi_2)^3
\eea
 In section 2 we discuss standard field theory in $1+1$ dimensions at finite
temperature.  In section 3 we derive the CSP formulation of field theory in a
box in $1+1$ dimensions, working in analogy with the standard 
finite temperature
formulation outlined in section 2.  We calculate the self energy using this CSP
formalism.   In section 4 we calculate the self energy using the box technique.
In section 5 we compare the results of the two calculations, and discuss our
conclusions.

\section{Field Theory in $1+1$ Dimensions}

In the next two sections we consider a non-interacting scalar field of mass m.

\subsection{Quantization}

We start from the Klein--Gordon equation of motion,
\bea
(\partial_t^2 -\partial_x^2 + m^2)\phi &=& 0 \label{KG1}\eea
The solution of the Klein--Gordon equation has the form
\bea
\phi \sim e^{i(\omega_k t \pm k x)};~~~~~\omega_k = \sqrt{k^2 + m^2} \nonumber
\eea
and describes the propagation of a plane wave of energy $\omega$. This equation
of motion should be obtainable from a least action principle.  If the action is
defined in terms of a Lagrange density as,
\bea
I=\int {\cal L} dt \nonumber
\eea then the equation of motion is determined from
\bea \delta I = 0 ~~~ \rightarrow~~~ \frac{\partial}{\partial t} \frac{\partial
{\cal L}}{\partial (\partial_t \phi)} - \frac{\partial {\cal L}}{\partial \phi}
=0 \nonumber\eea
which allows us to identify,
\bea
{\cal L} =   \frac{1}{2}[(\partial_t \phi)^2 - (\partial_x \phi)^2 - m^2 \phi^2
]\nonumber
\eea

To quantize the system we go to the Hamiltonian formalism.
We define canonical momenta
\bea {\Pi} = \frac{\partial{\cal L}}{\partial(\partial_t \phi)}= \partial_t
\phi \label{PI}
\eea and take the Legendre transform,
\bea
{\cal H} &=& {\Pi}\partial_x \phi - {\cal L}\xx
&=& \frac{1}{2} [(\partial_t \phi)^2 + \Pi^2 + m^2\phi^2] \label{Ham1}
\eea
Obtaining Hamiltonian's equations in the usual way we get,
\bea
\partial_t \phi = \frac{\partial {\cal H}}{\partial
{\Pi}};\,\,\,\,\,\,\,\partial_t {\Pi} = -\frac{\partial {\cal H}}{\partial
\phi}\nonumber
\eea
We can write these results as Poisson Brackets:
\bea
\partial_t\phi = \{\phi,{\cal H}\};\,\,\,\,\,\partial_t {\Pi} = \{ {\Pi},{{\cal
H}}\}\nonumber
\eea
To quantize we supplement
the initial conditions by the equal time commutation relations (ETCRs)
\bea
[\phi(x,t), \Pi(x',t)] = i\delta(x-x')\label{ETCR}
\eea
and replace the Poisson brackets with commutation relations,
\bea
i\partial_t \phi = [\phi,{H}];\,\,\,\,\,\,i\partial_t {\Pi} =
[{\Pi},{H}];~~~~~H(t) = \int dx {\cal H}(x,t)\label{Heqn}\nonumber
\eea

\subsection{Symmetries}
The energy momentum tensor,
\bea
\Theta_{\mu\nu} = \frac{\partial {\cal L}}{\partial (\partial^\nu \phi)}
\partial_\mu \phi - g_{\mu\nu} {\cal L}\label{EMT}
\eea
has for its zero-zero component: \bea
\Theta_{00} = {\cal H}\nonumber \eea
 From the conservation equation $\partial^\mu \Theta_{\mu\nu} = 0$
we obtain,
\bea
\frac{d}{dt}\int dx {\cal H} = \frac{d}{dt} H = 0\nonumber
\eea which means that the eigenvalues of $H$ are conserved quantities, and the
eigenstates of $H$ are stationary states.  Note that~(\ref{Heqn}) has solution
\bea
\phi(x,t) = e^{i{H}t}\phi(x,0) e^{-i{H}t};\,\,\,\,\,\,\Pi(x,t) =
e^{i{H}t}\Pi(x,0) e^{-i{H}t},\nonumber
\eea
which identifies ${H}$ as the generator of translations in time.

\subsection{Propagators}
Next we consider momentum-space expansions of the field variables.  We use the
stationary eigenstates of the Hamiltonian as basis states and expand around
solutions of the Klein--Gordon equation~(\ref{KG1}) by writing,
\bea
\phi(x,t) = \int \frac{dk_0}{2\pi(2\omega_k)}[a(k)e^{-i(\omega_k t-k x)} +
a^\dagger (k)e^{i(\omega_k t -k x)}]
\eea
where $\omega_k = \sqrt{k^2+m^2}$.
The creation and annihilation operators satisfy the commutation relation,
\bea
[a(k),a^\dagger (p)] = 2\pi (2\omega_k)\delta(k-p)
\eea
as a consequence of~({\ref{PI}) and~(\ref{ETCR}).  The interpretation of this
expression is as follows.  The frequency $\omega_k$ is always positive.  The
operator $a^\dagger(k)$ creates a mode of energy $\omega_k$ which propagates
forward in time, and the operator $a(k)$ destroys a mode of energy $\omega_k$,
which is equivalent to creating a mode with negative energy that propagates
backwards in time (an anti-particle).  Thus, by expanding in this set of
creation and annihilation operators we use a basis set that separates positive
and negative energy eigenstates.  This basis choice allows us to define time
causal propagators,
\bea
&&iD_R(x,t;y,t') = \theta(t-t') [\phi(x,t),\phi(y,t')] \nonumber \\
&&D_R(K) = \frac{1}{(k_0+i\epsilon)^2 - k^2 - m^2} = \frac{1}{(k_0 +
i\epsilon)^2 - \omega_k^2} \label{DR1} \eea
where translation invariance has been used in taking the Fourier transform.

\subsection{Finite Temperature Field Theory}
Finite temperature field theory involves expectation values of operators
weighted with Boltzmann factors,
\bea \langle {\cal O}\rangle = \sum_n \langle \phi_n | e^{-\beta H} {\cal O} |
\phi_n\rangle \nonumber\eea
The Boltzmann factor is equivalent to an imaginary time evolution operator
which gives,
\bea \langle {\cal O}\rangle = \sum_n\langle \phi_n(x);t-i\beta|{\cal
O}|\phi_n(x);t\rangle \nonumber \eea
We impose periodic boundary conditions in imaginary time,
\bea \phi_n(t-i\beta) = \phi_n(t)\nonumber
\eea
This constraint leads to Feynman rules which are the same as the zero
temperature rules except that the zero momentum components take discrete values
$k_0 = 2\pi i nT$, and the part of the loop integral that corresponds to
integrals over the zeroth momentum components are replaced by summations. In
the evaluation of Feynman integrals, these sums are replaced by contour
integrals that contain distribution functions which have poles at the same
discrete values $k_0 = 2\pi i n T$.  These contour integrals can be rotated so
that the integration contour encloses the real axis, running from negative
infinity to positive infinity above the real axis (the branch ${\cal C}_1$),
and backwards from positive infinity to negative infinity below the real axis
(the branch ${\cal C}_2$).  This contour is called the closed time path (CTP)
contour. The propagator can be written as a $2\times 2$ matrix of the form,
\bea
G=
\left(
\begin{array}{cc}
G_{11} & G_{12} \\
G_{21} & G_{22}
\end{array}
\right)\label{matrix1}
\eea
where $G_{11}$ is the propagator for fields moving along ${\cal C}_1$, $G_{12}$
is the propagator for fields moving from ${\cal C}_1$ to ${\cal C}_2$, etc.
The four components are given by,
\bea
&& G_{11}(X-Y) = -i\langle T \phi(X) \phi(Y))\rangle \nonumber \\
&& G_{12}(X-Y) = -i\langle \phi(Y) \phi(X) \rangle\nonumber \\
&& G_{21}(X-Y) = -i\langle \phi(X) \phi(Y) \rangle \nonumber \\
&& G_{22}(X-Y) = -i\langle \tilde{T}(\phi(X) \phi(Y))\rangle \label{G111}
\eea
where $T$ is the usual time ordering operator and $\tilde{T}$ is the
anti-chronological time ordering operator.
These four components are related by the condition,
\bea
G_{11} + G_{12} + G_{21} + G_{22} = 0 \label{constraint1}\eea
and therefore only three are independent.  We define the ``physical''
 combinations,
\bea
&&G_R = G_{11} - G_{12} =
-i\theta(t-t')\langle[\phi(x,t),\phi(y,t')]\rangle\nonumber \\
&&G_A = G_{11} - G_{21} =
i\theta(t'-t)\langle[\phi(x,t),\phi(y,t')]\rangle\nonumber \\
&&G_F = G_{11} + G_{22} =
-i\langle\{\phi(x,t),\phi(y,t')\}\rangle.\label{physical1} \eea
In momentum space the cyclicity of the trace allows us to obtain the
relationship between these propagators known as the KMS condition,
\bea
&&G_F(P) = N(p_0) (G_R(P) - G_A(P)) \label{KMS1} \\
&&N(p_0) = 1+2n(p_0)\,;~~~~n(p_0) = \frac{1}{e^{\beta p_0}-1}. \nonumber \eea

\section{The Closed Space Path}

\subsection{Formulation of the theory}

We start with a Klein--Gordon equation of the form,
\bea
(\partial_t^2 - \partial_x^2 - m^2)\phi(x,t) = 0 \label{KG2}\eea
which has solutions,
\bea
\phi\sim e^{i(k_0t\pm \omega_{k_0} x)}\,;~~~~\tilde{\omega}_{k_0} = \sqrt{k_0^2 +
m^2}\nonumber
\eea that describe the propagation of a plane wave with momentum
$\omega_{k_0}$. We should be able to obtain the equation of motion from a least
action principle of the form,
\bea
\tilde{I} = \int \tilde{\cal L}\,dx\,;~~~~ \delta \tilde{I} =
0\,~~~~\rightarrow ~~~~\frac{\partial}{\partial t} \frac{\partial \tilde{\cal
L}}{\partial (\partial_t \phi)} - \frac{\partial \tilde{\cal L}}{\partial \phi}
=0 \nonumber\eea
The Lagrangian that produces the correct equation of motion is,
\bea
\tilde{L} = \frac{1}{2}[\partial_x \phi)^2 - (\partial_t \phi)^2 - m^2\phi^2]
\eea

To obtain the Hamiltonian we define canonical momenta
\bea \tilde{\Pi} = \frac{\partial{\cal L}}{\partial(\partial_x \phi)} =
\partial_x \phi \nonumber
\eea
and take the Legendre transform,
\bea
\tilde{{\cal H}} = \tilde{\Pi}\partial_x \phi - {\cal L} =
\frac{1}{2}[(\partial_0 \phi)^2 + (\tilde{\Pi})^2 + m^2 ]\eea
Obtaining Hamiltonian's equations in the usual way we get,
\bea
\partial_x \phi = \frac{\partial \tilde{H}}{\partial
\tilde{\Pi}};\,\,\,\,-\partial_x \tilde{\Pi} = \frac{\partial
\tilde{H}}{\partial \phi}\nonumber
\eea
We can write these results as Poisson Brackets:
\bea
\partial_x\phi = \{\phi,\tilde{\cal  H}\};\,\,\,\,\,\partial_x \tilde{\Pi} = \{
\tilde{\Pi},\tilde{{ \cal H}}\}\nonumber
\eea
To quantize we supplement the initial conditions with the equal place
commutation relations (EPCRs)
\bea
[\phi(x,t), \tilde{\Pi}(x,t')] = i\delta(t-t')\label{EPCR}
\eea
and replace the Poisson brackets with commutation relations,
\bea
i\partial_x \phi = [\phi,\tilde{H}]\,;\,\,\,\,\,\,i\partial_x \tilde{\Pi} =
[\tilde{\Pi},\tilde{H}]\,;~~~~ \tilde{H}(x) = \int dt {\tilde{\cal
H}}(x,t)\label{Heqn2}.
\eea
The energy momentum tensor is obtained from~(\ref{EMT}) as before.  The 1--1
component is $\Theta_{11} = -\tilde{H}$ and the constraint
$\partial^\mu\theta_{\mu\nu}=0$ leads to
\bea
\frac{d}{dx}\int\,dt\,\tilde{H} = 0 \nonumber
\eea
which means that the eigenvalues of $\tilde{H}$ are conserved and that the
eigenstates are stationary states.
Note that~(\ref{Heqn2}) has solutions,
\bea
\phi(x) = e^{i\tilde{H}x}\phi(0) e^{-i\tilde{H} x};\,\,\,\,\,\,\Pi(x) =
e^{i\tilde{H}x}\Pi(0) e^{-i\tilde{H} x}\nonumber
\eea
which identifies $\tilde{H}$ as the generator of translations in space.

To obtain an expression for the propagators, we use a basis set of the
stationary eigenstates of the Hamiltonian $\tilde{H}$ and expand around
solutions of the equation of motion~(\ref{KG2}) by writing,
\bea
\phi(x,t) = \int \frac{dk_0}{2\pi(2\tww)}[a(k_0)e^{-i(\tww x-k_0 t)} +
a^\dagger (k_0)e^{i(\tww x - k_0 t)}].\label{EXP2}
\eea
The creation and annihilation operators satisfy the commutation relation,
\bea
[a(k_0),a^\dagger (p_0)] = 2\pi (2\tww)\delta(k_0-p_0)\label{EPCR2}
\eea
The interpretation of this expansion is as follows. The operator
$a^\dagger(k)$ creates a mode of momentum $\tilde{\omega}_{k_0}$ which propagates
forward in space, and the operator $a(k)$ destroys a mode of momentum
$\tilde{\omega}_{k_0}$, which is equivalent to creating a mode with negative momentum
that propagates backwards in space.
By expanding in momentum eigenstates, instead of energy eigenstates, we have
obtained a formalism in which it is simple to obtain propagators that are
retarded in space.

The space-retarded propagator is defined as,
\bea
iD_{R}(x,t;y,t') =  D_{11}  - D_{12} = \theta(x-y)[\phi(x,t),\phi(y,t')]
\nonumber
\eea
Substituting in the expressions for the momentum expansions of the
fields~({\ref{EXP2}) and using the EPCR's~(\ref{EPCR2}) we obtain, in momentum
space,
\bea D_R(K) = -\frac{1}{k_0^2 - (k+i\epsilon)^2 + m^2} =
\frac{1}{(k+i\epsilon)^2 - \tww^2} \label{DR2}
\eea
which can be compared to the result for the time-retarded
propagator~(\ref{DR1}).

We can write the propagator in matrix form on the CSP contour.  The result is
analogous to the CTP form~(\ref{matrix1}),
\bea
D=
\left(
\begin{array}{cc}
D_{11} & D_{12} \\
D_{21} & D_{22}
\end{array}
\right)\label{matrix2}
\eea
with (compare(~\ref{G111})),
\bea
&& D_{11}(X-Y) = -i\langle S \phi(X) \phi(Y))\rangle \nonumber \\
&& D_{12}(X-Y) = -i\langle \phi(Y) \phi(X) \rangle\nonumber \\
&& D_{21}(X-Y) = -i\langle \phi(X) \phi(Y) \rangle \nonumber \\
&& D_{22}(X-Y) = -i\langle \tilde{S}(\phi(X) \phi(Y)).\rangle \label{G112}
\eea
The time ordering operator is replaced with a space ordering operator $S$, and
$\tilde{S}$ is the reverse space ordering operator.  The four components are
related by the condition (compare~(\ref{constraint1})),
\bea
D_{11} + D_{12} + D_{21} + D_{22} = 0 \label{constraint2}\eea
 We define the ``physical'' combinations (compare~(\ref{physical1})),
\bea
&&D_R = D_{11} - D_{12} =
-i\theta(x-x')\langle[\phi(x,t),\phi(y,t')]\rangle\nonumber \\
&&D_A = D_{11} - D_{21} =
i\theta(x'-x)\langle[\phi(x,t),\phi(y,t')]\rangle\nonumber \\
&&D_F = D_{11} + D_{22} =
-i\langle\{\phi(x,t),\phi(y,t')\}\rangle.\label{physical2} \eea
The KMS--like condition is obtained by analogy with~(\ref{KMS1}),
\bea
&&D_F(P) = N(p)(D_R(P) - D_A(P)) \label{KMS2} \\
&&N(p)={\rm cot}(\frac{1}{2}L p) = i(1+2n(p));~~~~~n(p) = \frac{1}{e^{iLp}-1}
\nonumber
\eea

\subsection{Calculation of the imaginary part of the retarded self energy}

We consider the contribution to the self energy of the field $\phi_1$ arising
from a loop with one $\phi_1$ propagator and one $\phi_2$ propagator on
internal lines.  Other contributions to the $\phi_1$ self energy can be
obtained in a straightforward way by setting $m_1=m_2$. The spatially retarded
self energy is given by,
\bea
\Sigma_R(P)~&& = ~~\Sigma(P)_{11} + \Sigma(P)_{12} \nonumber \\
&& = i\lambda^2 \int dK (D_{11}(K) D_{11}(K+P) - D_{21}(K) D_{12}(K+P)
)\nonumber \\
&&= i\frac{\lambda^2}{2}\int dK (D_F(K) D_R(P+K) + D_A(K) D_F(P+K))
\eea
where $dK = dk_0\,dk\,/(2\pi)^2$, $D_R$ is given by~(\ref{DR2}), and $D_A$ is
the complex conjugate of $D_R$. We use the KMS condition~(\ref{KMS2}) to write,
\bea
D_F(K) = N(k) 2i {\rm Im} D_R(K) = -2i\pi N(k) \epsilon(k) \delta(k^2 -
\tilde{\omega}_{k_0}^2)\nonumber
\eea
Using the delta functions to do the $k$ integrals we obtain,
\bea
\Sigma_R(P) = \frac{\lambda^2}{4\pi}{\cal E}_{m_1,m_2} \int
\frac{dk_0}{\tww}\,N(\tilde{\omega}_{k_0})
\left(\frac{1}{(p+\tww + i\epsilon)^2 - (\twwp)^2} +
\frac{1}{(p-\tww +i\epsilon)^2 - (\twwp)^2 }\right)\nonumber
\eea
where ${\cal E}_{m_1,m_2} f(m_1,m_2) = f(m_1,m_2) + f(m_2,m_1)$.
To get the imaginary part, we take the imaginary part of the two fractions
using
\bea
i{\rm Im}\left(\frac{1}{(\alpha +i\epsilon)^2 - \beta^2}\right) = -i\pi
\epsilon(\alpha) \delta(\alpha^2 - \beta^2) \nonumber
\eea
The two terms can be combined using the symmetrization factor 
${\cal O}_p f(p)= f(p) - f(-p)$.
The result is,
\bea
i{\rm Im}\Sigma_R(P) = -i\frac{\lambda^2}{8}{\cal E}_{m_1,m_2} {\cal O}_p \int
dk_0\,N(\tww)\frac{1}{\tww} \epsilon(p+\tww) \delta [(\tilde{\omega}_{k_0} +
p)^2 - (\tilde{\omega}'_{k_0+p_0})^2]  \label{xx5}
\eea
We define,
\bea
&& x = P^2-\delta m^2 + 2p_0k_0 \label{xx6} \\
&& \alpha = 2p\tww \nonumber
\eea
which allows us to write,
\bea
\delta[(\tilde{\omega}_{k_0} + p)^2 - (\tilde{\omega}'_{k_0+p_0})^2]
\label{delta}
= \delta(x-\alpha).
\eea
We rewrite the delta function using,
\bea
\delta(x-\alpha) = 2\alpha [\theta(x)\theta(\alpha) -
\theta(-x)\theta(-\alpha)]\delta(x^2-\alpha^2) \nonumber
\eea
We obtain,
\bea
&&x^2-\alpha^2 = 4P^2(k_0-k_{01})(k_0-k_{02})\nonumber \\
&&k_{01,2} = \frac{1}{2P^2}(-p_0(P^2 - \delta m^2) \pm  p \sqrt{R})
\label{ks}\\
&&R = (P^2+(\Sigma m)^2)(P^2+(\delta m)^2) \nonumber \\
&&\Sigma m = m_1+m_2;~~~~~\delta m = m_1-m_2;~~~~~P^2 = p_0^2-p^2. \nonumber
\eea
from which we derive,
\bea
\delta(x-\alpha) = \frac{1}{\sqrt{R}}\tilde{\omega}_{k_0} \theta(xp)
(\delta(k_0-k_{01}) + \delta(k_0-k_{02}))
\nonumber
\eea
Substituting into~(\ref{xx5}) we obtain,
\bea i {\rm Im}\Sigma_R(p_0,p) = i\frac{\lambda^2}{8\sqrt{R}} {\cal
E}_{m_1,m_2} {\cal O}_p \int dk_0 N(\tilde{\omega}_{k_0})
\epsilon(p+\tilde{\omega}_{k_0}) \theta(xp) (\delta(k_0-k_{01}) +
\delta(k_0-k_{02})) \label{22}
\eea
We define,
\bea
a_{1,2} = \frac{1}{P^2} (p_0\delta m^2 \pm p \sqrt{R}) \label{as}
\eea
which allows us to write,
\bea
&&k_{01,2} = \frac{1}{2}(-p_0 + a_{1,2}) \nonumber\\
&&\frac{x(k_{01,2})}{p} = \frac{x_{1,2}}{p} = -p+a_{1,2} \nonumber \\
&&\omega_{k_01,2} = \omega_{1,2} = \frac{1}{2p^2}|px_{1,2}| \nonumber \\
&&\epsilon(p+\omega_{1,2}) = \epsilon(p+a_{1,2}) \nonumber
\eea
Substituting these results into~(\ref{22}) we obtain,
\bea
 {\rm Im}\Sigma_R(p_0,p) &&= -i\frac{\lambda^2}{8\sqrt{R}} {\cal E}_{m_1,m_2}
{\cal O}_p \nonumber \\
&&~~[N(\frac{1}{p}(-p + a_1)) \theta(-p + a_1)\epsilon(p + a_1) +
N(\frac{1}{p}(-p + a_2)) \theta(-p + a_2)\epsilon(p + a_2)]
\eea
To simplify this result we expand out the symmetrization factor ${\cal O}_p$
using,
\bea p\rightarrow -p ~~~~\Rightarrow~~~~ a_{1,2} \rightarrow -a_{2,1} \nonumber
\eea
The result is,
\bea
i {\rm Im}\Sigma_R(p_0,p) =-i\frac{\lambda^2}{8\sqrt{R}} {\cal E}_{m_1,m_2}
[N(p-a_1) \epsilon(p^2-a_1^2) + N(p-a_2) \epsilon(p^2-a_2^2)] \label{xx4}
\eea

\section{The Box Calculation}

In this section we calculate the same contribution to the self energy using the
standard Feynman rules of zero temperature field theory, and explicitly
constrain the system to a box of length $L$.  The self energy is given by,
\bea
\Sigma(P) =i\lambda^2 \int dK
\frac{1}{[K^2-m_1^2+i\epsilon]}\frac{1}{[(K+P)^2-m_2^2+i\epsilon]}.
\eea
 We restrict the variable $k$ to values $k=2\pi n/L$ where $L$ is the length of
the box, and $n$ is an integer.  The integral becomes,
\bea
\Sigma(p_0, p_m) =i\lambda^2 \int dk_0 \frac{1}{L} \sum_n \left( \frac{1}{k_0^2
- k_n^2-m_1^2+i\epsilon}\right) \left( \frac{1}{(k_0+p_0)^2 - (k_n+
p_m)^2-m_2^2 +i\epsilon}\right)
\eea
We can replace the sum over the discrete index by a contour integration:
\bea
\sum_n f(p=\frac{2\pi n}{L}) = \frac{1}{2\pi i}\frac{L}{2} \int dp \,\,{\rm
cot}(\frac{1}{2}LP) f(p)
\eea where the contour is a clockwise loop that encloses the poles of the
cotangent function which lie along the real axis.
This gives,
\bea \Sigma(p_0,p_m) = -i\frac{\lambda^2}{2} \int \frac{dk_0}{2\pi} \int
\frac{dk}{2\pi i} \,\,{\rm cot} (\frac{1}{2}Lk) \left( \frac{1}{k_0^2 -
k^2-m_1^2+i\epsilon}\right) \left( \frac{1}{(k_0+p_0)^2 -(k+
p_m)^2-m_2^2+i\epsilon}\right)\nonumber
\eea
Using~(\ref{KMS2}) we rewrite the cotangent function,
\bea
{\rm cot}(\frac{1}{2}Lp)= i(1+2n(p)) = N(p) \label{N1}
\eea

We deform the contour to two semi-circles in the upper and lower half planes.
We do the $k$ integral first, picking up contributions from four poles.  These
four terms can be combined using the two factors ${\cal E}_p$ and ${\cal
E}_{m_1,m_2}$ where,
${\cal E}_p f(p) = f(p) + f(-p)$.
We use the notation $\omega_{k_0} = \sqrt{k_0^2 - m_1^2}$, $\omega'_{k_0} =
\sqrt{k_0^2 - m_2^2}$ etc. Note that these frequencies can be imaginary.  When
doing the contour integral, it doesn't matter if the frequencies are real or
imaginary, since both halves of the contour form counter clockwise loops, and
thus pole contributions from both sides of the real axis have the same sign.
The result is,
\bea \Sigma(p_0, p_m) = \frac{\lambda^2}{8\pi} {\cal E}_{m_1,m_2} {\cal E}_p
\int dk_0\,N(\omega_{k_0}) \frac{1}{\omega_{k_0}} \frac{1}{(\omega_{k_0} + p_m)^2
- \omega'^2_{k_0+p_0} -i\epsilon} \nonumber
\eea
To find the imaginary part of the spatially retarded piece we follow
the analogous procedure as that used for finite temperature
field theory in the imaginary time formalism: we analytically
continue $p_m$ to the whole complex $p$ plane and take the discontinuity,
\begin{eqnarray}
 i{\rm Im}\Sigma_R(p_0,p)&&= \frac{1}{2}(\Sigma(p+i\epsilon) -
\Sigma(p-i\epsilon)) \nonumber \\
&&= -i\frac{\lambda^2}{8} {\cal E}_{m_1,m_2} {\cal O}_p \int dk_0
\,N(\omega_{k_0}) \frac{1}{\omega_{k_0}} \epsilon(p+\omega_{k_0})
\delta[(\omega_{k_0} + p)^2 - \omega'^2_{k_0+p_0}]
\label{11}
\end{eqnarray}
Comparison of~(\ref{11}) with~(\ref{xx5}) shows immediately that
\bea ({\rm Im}\Sigma_R(m^2))_{Box} = ({\rm Im}\Sigma_R(-m^2))_{CSP}
\eea
Note that this result is what we would have expected from~(\ref{KG2}) in which
the mass term has the opposite sign relative to~(\ref{KG1}).

\section{Conclusions}

The result for the imaginary part of the spatially retarded self energy
contains a cotangent factor.  This factor is a direct consequence of the fact
that the system has been constrained to a box, and gives rise to the $1+1$
dimensional analogue of the Rabi oscillations that have been calculated
non-relativistically from the Jaynes-Cummings model.  In our case the
oscillations are given by a cotangent function, instead of a sine or cosine
function, and thus are discontinuous.  It is expected that this behaviour is an
artifact of working in $1+1$ dimensions.

We have shown that it is possible to obtain spatially retarded green functions
for a system confined to a box by using a formulation of field theory that is
similar to the  closed time path formulation of finite temperature field
theory. All of the powerful machinery of the CTP method can be taken over when
calculating in a one dimensional box by using the CSP formalism derived in this
paper. We have calculated the self energy in a simple scalar theory as an
example, and verified that the result is related in a straightforward way to
the result that would be obtained in the traditional `box' calculation.  The
possible advantages of our method are the same as those of the CTP
method of finite temperature field theory. In particular,
it is typically easier to generalize to higher $n$-point
functions using such an approach.  When calculating in the traditional manner, it is necessary to do
an analytic continuation from discrete momentum values to the full complex
plane, and this analytic continuation becomes extremely complicated for higher
n-point functions.

\end{document}